# A Real-time Localization System Using RFID for Visually Impaired

Qinghui T.[1], Malik M.Y.[2], Youngjee H.[1], Jinwoo P.[1]

**Dept. of Industrial Engineering and ASRI[1], Dept. of EECS[2]**

**Seoul National University, Seoul, Korea**

## Abstract

Gadgets helping the disabled, especially blind that are in least accessibility of information, use acoustic methods that can cause stress to ear and infringe user's privacy. Even if some project uses embedded Radio Frequency Identification (RFID) into the sidewalk for blind's free walking, the tag memory design is not specified for buildings and road conditions. This paper suggested allocation scheme of RFID tag referring to EPCglobal SGLN, tactile method for conveying information, and use of lithium battery as power source with solar cells as an alternative. Results have shown independent mobility, accidents prevention, stress relief and satisfied factors in terms of cost and human usability.

## 1. Introduction

This is the era of information technology and everyone has access to information. Lives of human beings are facilitated and they can reaped the benefits of cutting edge (overuse the same words). However, we compare the technical resources available for visually impaired, still 'digital divide' between their lives and that of the others can be sensed. Specifically, for the blind's walking independently on the road, many researchers have suggested the white canes with audio system and it is most state of the art. But continuously hearing sound output or wearing a headset can be irritating to the system users. Therefore, this research proposed a serialized RFID guide system in real time. We referred to the EPCglobal and suggested information allocation scheme of RFID tag. EPCglobal is a global standards system that combines RFID technology, existing communications network infrastructure, and the Electronic Product Code [1]. EPCglobal is to facilitate the exchange of information and object between trading partners in supply chain, so the identification, data capture, and data exchange are the needs. However our suggestion is related to service engineering of social infrastructure, we focus on just identification and data capture with offline concept. Especially information elements of SGLN (Serialized Global Location Number) give a hint in tag scheme. Users could get 'where I am' from the information of location, roads, and nearby buildings in tactile way, therefore confidently mobility, avoiding ear stress, and guaranteeing privacy are given.

## 2. Related Works

D'Atri, E. et al. introduced a RFID cane reader and PDA system. PDA communicated with the RFID cane for location information and user has to carry PDA [2]. Ortigosa, N. et al. presented mobility assistance aimed to sonicate to the user the presence of obstacles and free path by using an optical laser distance meter, a stereo-camera and GPS system [3]. Another localization system for blind using GPS network is introduced by S. Chumkamon et al [4]. Main focus of their work is in helping the blind to get know of their location in indoor environments. Yuriko Shiizu et al proposed an indoor guidance arrangement relying on different colors of lines for navigation [5]. This scheme also incorporates a smart cane in which the sensors for color recognition are embedded. Guidance voice allows the user to know about his location in this proposed scheme. For these proposals, user always has to carry some devices along with the white cane and information is transmitted vocally. Additionally, power resources are very limited in these systems.

## 3. Proposed System Configuration

In our paper we considered the use of tactile signals for giving information rather than acoustic signals, and contributed better power source for the operation of the system. Our system is based on RFID passive rags, reader, control unit, Braille display and lithium and solar power source. A reader incorporated in white cane reads the information from the passive tags, then retrieves data from memory and transfers this information to the Braille display so that the visually impaired can read it. Passive RFID tags can be installed on roads, Braille blocks, street lamps, buildings or other signs and location indication boards. Reader can read the information from tag, analyze the data and pass the information to the control unit. Control unit helps in presenting the useful information to the blind people through Braille display.

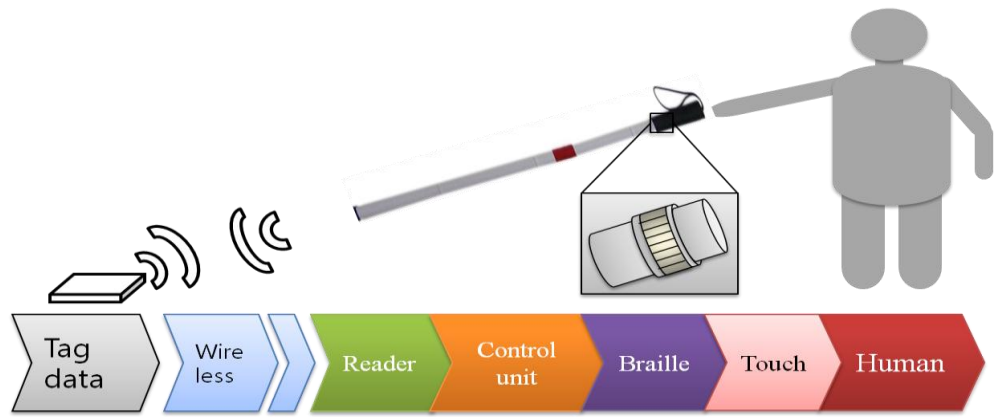

Figure 1: System configuration

## 4. RFID System

EPC Global develops industry-driven standards for the Electronic Product Code™ (EPC). Most of the tag formations are being designed using EPC Global GEN2 standard. These EPC tag forms are simple and inexpensive.

Our proposed system is cost effective and easy to implement because of use of this standard to define our RFID tags.

**4.1. RFID Tag Definition**

This research used EPC Global Serialized Global Location Number-96 (SGLN-96) formation. SGLN-96 contains header, filter value, partition, company prefix, location reference, and extension component fields [6].

Table 1: Company prefix, location reference and extension component allocation

| 1. Company Prefix (20 bits) | | 2. Location Reference (21 bits) | | | 3. Extension Component(41bits) | | | |
|---|---|---|---|---|---|---|---|---|
| Company code | Direction code | Main road | Sub road | Path | Building number | Direction | Road condition | Serial number |
| 18 bits | 2 bits | 6 bits | 7 bits | 8 bits | 10 bits | 2 bits | 5 bits | 23 bits |

As shown in Table 1, we divided company prefix into two parts; company code and direction code. We can define nearly 260 thousand building names using 18 bits of company code. 2 bits of direction code are used to represent relative direction of tag and building. The code values of right and left directions are 00 and 11 respectively, and that of forward and backward direction are 10 and 01. This formation will facilitate figuring of direction in the protocol. Generally, the address definition of building is based on the road layouts. We also used this method to standardize location reference code in SGLN-96. We categorized roads into three types: main road, sub road and path. 6, 7 and 8 bits are respectively allotted to these road categories. Up to 2 million pathways can be defined in the tag code.

In extension component part, the first bit's value is 1. We allocated next 10 bits to building numbers. Building numbers are allotted from the start of the road, with odd numbers assigned to the left side and even numbers assigned to the right side, as shown in Figure 2. We used 2 bits to indicate the direction of road conditions and the other 5 bits to mark that road condition. Road conditions may refer to several detailed information to help blind pedestrians, like building entrance, stairs, pedestrian crossings, traffic light, turn etc. Remnant 23 bits are used for defining passive tag serial number. These serial numbers increase with respect to the right-hand side of the road. [4]and assume this direction to be positive,

**4.2. RFID Tag Layout Design**

Figure 2 shows a general RFID tag infrastructure layout. The passive RFID tags covered by the protective shield are attached to the sidewalks. When the tag infrastructure is constructed, the tag serial number increases from the start of road to the end which is attached in the road. Although the distance between two tags should be according to the road conditions, we proposed it to be near 8 m. In case of dense road conditions, tags can be positioned closer in order to provide the blind with more information.

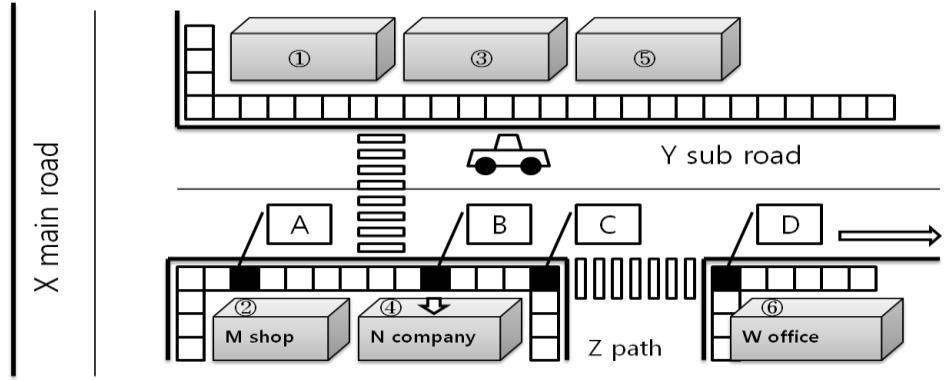

Figure 2: Real road condition and tag layout

### 4.3. RFID Tag Analysis Protocol

Now we will illustrate our suggested tag protocol. Information of the four tags shown in Figure 2 is shown in Table 2. When a pedestrian walks on Y sub road from left-hand side to right-hand side, the tags are read sequentially and data from them is stored in memory stack in reader. The received data comply with FIFO (first in, first out) rule in the queue. The simplified data analysis algorithm is shown in Figure 3.

Table 2: Tag sample

| Tag | Tag code (0 = Null) | | | | | | | | | |
|---|---|---|---|---|---|---|---|---|---|---|
| A | M shop | right | 0 | Y sub road | 0 | No 2 | forward | 0 | 0030 | M shop |
| B | N company | right | 0 | Y sub road | 0 | No 4 | right | entrance | 0060 | N company |
| C | N company | right | 0 | Y sub road | 0 | No 4 | forward | crosswalk | 0070 | N company |
| D | W office | right | 0 | Y sub road | 0 | No 6 | back | crosswalk | 0071 | W office |

When the reader came in contact with installed passive tag, the tag is verified and its data is stored in reader's memory stack. Tag's serial number is compared with the previous entry of stack. As defined earlier, tag serial number is increasing on the right side of the road. If the pedestrian is walking in the positive direction, serial numbers read by the reader increases in value. The reader retrieves information and sends message to the pedestrian. When the pedestrian walks from negative direction, the system can judge the pedestrian's direction by decrement in serial numbers. In this case, the system would convert the company prefix and extension component directions in Table 1 to their opposite values by NOT operation.

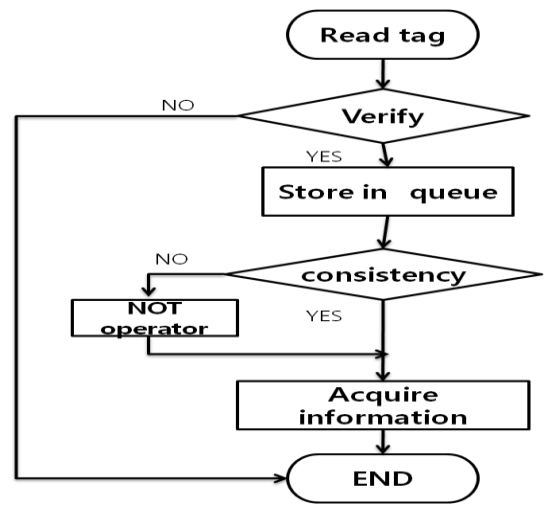

Figure 3: RFID tag analysis algorithm

It solved bidirectional confusion which can be found in previous systems. When tag B is registered by the reader moving from left-hand side to right-hand side, the pedestrian is given information about the entrance of N company and its direction (right-hand side). On the contrary, pedestrian is told that entrance of N company is on left-hand side.

## 5. Proposed Smart Cane

This paper proposed a smart cane, which is responsible for tag reading operation and communicating the data to the user. We have underlined some of the aspects that can enable effective design of the cane. As the smart white cane will perform two significant operations, there is a need of efficient power source for its operation.

### 5.1 RFID Reader

Class 1 (Gen 2) UHF based reader can be deployed in smart cane, because of its large range of reading. They are able to read passive tags within 10 m. Working at much higher frequencies will enable design of small antennas which can easily be embedded in pattern of the PCB. This will help in keeping the reader module size to minimum and can be easily incorporated in the white canes with minor modifications.

### 5.2 Information display part as Braille part

Smart cane is able to communicate the information received by the reader to the user. It uses Braille for transferring the information to the user. User can read the data presented by smart cane conveniently. Braille is the best known communication system for blind people, developed by Louise Braille. In this system characters are exhibited in a pattern of six (or more) dots having two columns of three dots, known as Braille cell. Such cell is shown in Figure 4. A combination of six dots can only have 64 different configurations. For many languages, basic communication can be performed by the use of six dot cells. There are many standards available for Braille cell dimensions, dot size, distance between lines and many other parameters.

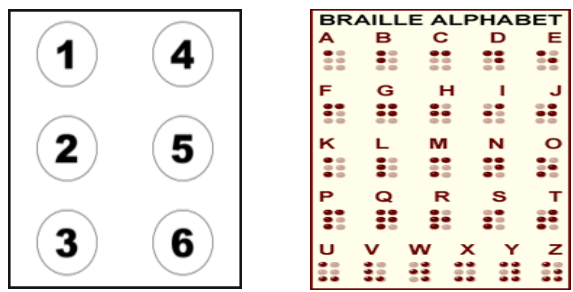

Figure 4: Braille cell and representation of English alphabets by Braille cells [8,9]

### 5.3 Braille Unit

For conveying information to the user, Braille display on smart cane uses two lines for displaying the characters. First line will give indication of nearby building, road, street or any other important location around. The second line will provide information about any specific place around that location, like the presence of any pedestrian crossing, community services (elevator, washrooms, etc) and telephone booth. Directions of these places are also shown on the second line.

The width of the smart cane at the handling side can be from 110-120 mm. This size is appropriate as it is easy to handle. In common Braille systems dot height is approximately 0.5 mm; the horizontal and vertical spacing between dot centers within a cell is approximately 2.5 mm; the blank space between dots on adjacent cells is 3.75 mm horizontally [10]. If we consider using the common Braille system for displaying data on smart cane nearly 12 characters can be easily shown on a single line, which are enough for conveying basic information. By using two lines we can easily inform the user about the location around him and places of importance and interest surrounding him.

### 5.4 Control unit between reader and Braille display

Our proposed Braille display works on selection of dots and electromechanical switching. Any dot in display system can be chosen to be raised to represent a dot or can be lowered to act as an empty place. In our system reader gets information about location from the passive tags installed and transfers it to a control unit. This control unit is responsible for the mechanism of displaying message on Braille display. The control unit consists of a microcontroller unit (MCU) which can be programmed to facilitate selection of dots using decoders and buffer based on the commands by the MCU. Electrical signals from decoders and buffers can operate electromechanical switches which can raise or lower metallic or plastic pins. By using this simple layout, we can represent data on our Braille display.

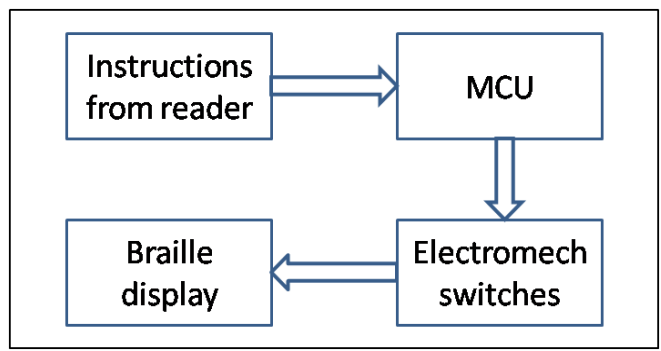

Figure 5: Outline of the system in smart cane

### 5.5 RF reader and Braille display power source

Choice of the power system for different units in smart cane is very crucial. Our system uses Lithium ion batteries (LIB) as the power source for the RF reader and control unit. They have best energy-to-weight ratios, no memory effect, and a slow loss of charge when not in use. LIB can be recharged easily which will help in minimizing the required maintenance. When the ubiquitous system of the smart cane is not being used, it can be switched off. The ubiquitous activity of smart cane will primarily be used outdoors, which makes it possible to utilize solar energy to charge another cell battery to function like uninterruptible power supply. When the LIB is near depletion, low battery alarm will be given to the user and reader and Braille display power source will be changed from LIB to solar cell battery. Like the cellular phone, LIB can be charged easily and conveniently.

## 6. Results

We evaluate our prototype comparing with previous guide systems based on three factors of context of use including acceptability, usability, and cost benefit. The acceptability tells whether the product will actually be used in real life. The usability represents the user-friendliness of a system. The cost benefit compares unit cost of previous research with ours [12]. For the acceptability factor, our system requires little provision of training to support initial usage and adaptive activities. Setting the RFID tag and power source inside cane, and attaching tactile brail and solar cell outside cane are only different from original white cane. For implementation, installing RFID on the road block, just power on and getting tactile messages are enough. These satisfy the usability factor. Most of the assistance prototypes are equipped with GPS, sensors, or portable navigation with headset. Those are normally twice expensive than ours just consisting of MCU, lithium battery, and tactile brail. These satisfy the cost benefit factor.

## 7. Conclusion and future work

This research presents tag design and smart white cane for ease in reading information without extra devices.
Short evaluation is done according to context of use which considers main user, task and environmental characteristics of the situation in which it will be operated [12]. The prototype is easy to learn, use, and buy or be subsidized from public service, so we concluded the acceptability, usability, and cost benefit are satisfied. Also huge decision opportunities and visibility for the blind in working are given. Future work will be network environment and compatibility with infrastructure and current standard. Extending network environment to system can enable range to be not only high-value applications of walking decisions, but also tracking risk situations which call for help through the network. Road or building condition changes could be easily updated by the server. Compatibility with existing infrastructure and current standard such as EPCglobal tag scheme are needed for smooth application.

# APPENDIX A

Table A1 shows the bit allocation in EPC SGLN-96 standard by EPCglobal.

Table A1: EPC SGLN-96 bit allocation

|         | Header | Filter Value | Partition | Company Prefix | Location Reference | Extension Component |
|---------|--------|--------------|-----------|----------------|--------------------|---------------------|
| SGLN-96 | 8      | 3            | 3         | 20-40          | 21-1               | 41                  |

Table A2 represents some of the Braille standards being used in the world.

Table A2: Some Braille standards

|                    | Horizontal dot to dot (mm) | Vertical dot to dot (mm) | Cell to cell (mm) | Line to line (mm) | Dot base diameter (mm) | Dot height (mm) |
|--------------------|----------------------------|--------------------------|-------------------|-------------------|------------------------|-----------------|
| Electronic Braille | 2.4                        | 2.4                      | 6.4               |                   |                        | 0.8             |
| French             | 2.5 - 2.6                  | 2.5 - 2.6                |                   | >10               | 1.2                    | 0.8 - 1.0       |
| German             | 2.5                        | 2.5                      | 6.0               | 10.0              | 1.3 - 1.6              | $\geq 0.5$      |
| Small English      | 2.03                       | 2.03                     | 5.38              | 8.46              | 1.4 - 1.5              | 0.33            |